\documentclass[journal]{IEEEtran}
\ifCLASSINFOpdf
\else
\fi
\usepackage{cite}
\usepackage[cmex10]{amsmath}
\usepackage{amssymb}
\usepackage{algorithmic}
\usepackage{array}
\usepackage{mdwmath}
\usepackage{booktabs}
\usepackage{mdwtab}
\usepackage{eqparbox}
\usepackage{graphicx}
\usepackage[justification=centering]{caption}
\usepackage{bm}
\usepackage{float}
\usepackage [latin1]{inputenc}
\begin{document}

\title{ECG Identification under 
 Exercise and Rest  Situations via Various Learning Methods}

\author{Zihan~Wang, Yaoguang~Li, and Wei~Cui~\IEEEmembership{Senior Member,~IEEE} 
\thanks{
This work was supported by the National Natural Science Foundation of China under Grant 61873317, and in part by the Fundamental Research Funds for the Central Universities.
 \textit{Corresponding author: Wei Cui.}

The authors are with the School of Automation Science and Engineering, South China University of Technology, Guangzhou 510641, China~(e-mail: aucuiwei@scut.edu.cn).
}}

\markboth{Submitted to IEEE Transactions on Neural Networks and Learning Systems}
{Shell \MakeLowercase{\textit{et al.}}: Bare Demo of IEEEtran.cls for Journals}
\maketitle
\IEEEpeerreviewmaketitle

\begin{abstract}
As the advancement of information security, human recognition as its core technology, has absorbed an increasing amount of attention in the past few years. A myriad of biometric features including fingerprint, face, iris, have been applied to security systems, which are occasionally considered vulnerable to forgery and spoofing attacks. Due to the difficulty of being fabricated, electrocardiogram (ECG) has attracted much attention. Though many works have shown the excellent human identification provided by ECG, most current ECG human identification (ECGID) researches only focus on rest situation. In this manuscript, we overcome the oversimplification of previous researches and evaluate the performance under both exercise and rest situations, especially the influence of exercise on ECGID. By applying various existing learning methods to our ECG dataset, we find that current methods which can well support the identification of individuals under rests, do not suffice to present satisfying ECGID performance under exercise situations, therefore exposing the deficiency of existing ECG identification methods.
\end{abstract}

\section{Introduction}
With the increasing demand for information security and reliability of protective measures, human identification has become an important research topic in corresponding field. However, traditional identification technologies such as certificates  and passwords are likely to be forgotten, copied and even stolen, which may cause harm to people's personal information, therefore no longer meet their needs for security. Due to the rising awareness of people and the rapid development of science and technology, biometric identification technology has emerged, which is based on the unique anatomical, physiological or behavioural characteristics hardly possible to be forged. So far, common biometric recognition methods \cite{Jain:2004,Pourbabaee:2018, Silipo:1998,Zheng:2019,Xu:2015,Zhang:2019} include face, fingerprint, voice, which are mature and have high recognition rates, but not perfect enough. For example, a FaceID can be corrupted by photograph or make-up; a fingerprint can be copied and recreated with latex; a voice can be recorded or imitated as well. In order to strengthen the reliability and security of biometric identification technology, on the one hand, some  scholars mix up multiple biometric technologies to make it harder for the system to be cracked. On the other hand, new reliable biometric identification techniques are being increasingly developed such as ECGID which, as one of the hottest research direction, has been developing for almost 20 years. The location, size and structure of the heart are so various from person to person that each person's ECG signal is regarded unique~\cite{Chan:2006}, and the difference of which provides a theoretical basis for the identification of ECG signals. Compared with the traditional biometric signal, ECG signals are bioelectrical signals generated by living bodies, making it more difficult to fake~\cite{Wang:2006,Agrafioti:2008,Agrafioti:2010}.

\begin{figure}
  \centering
  \includegraphics[width=8.5cm]{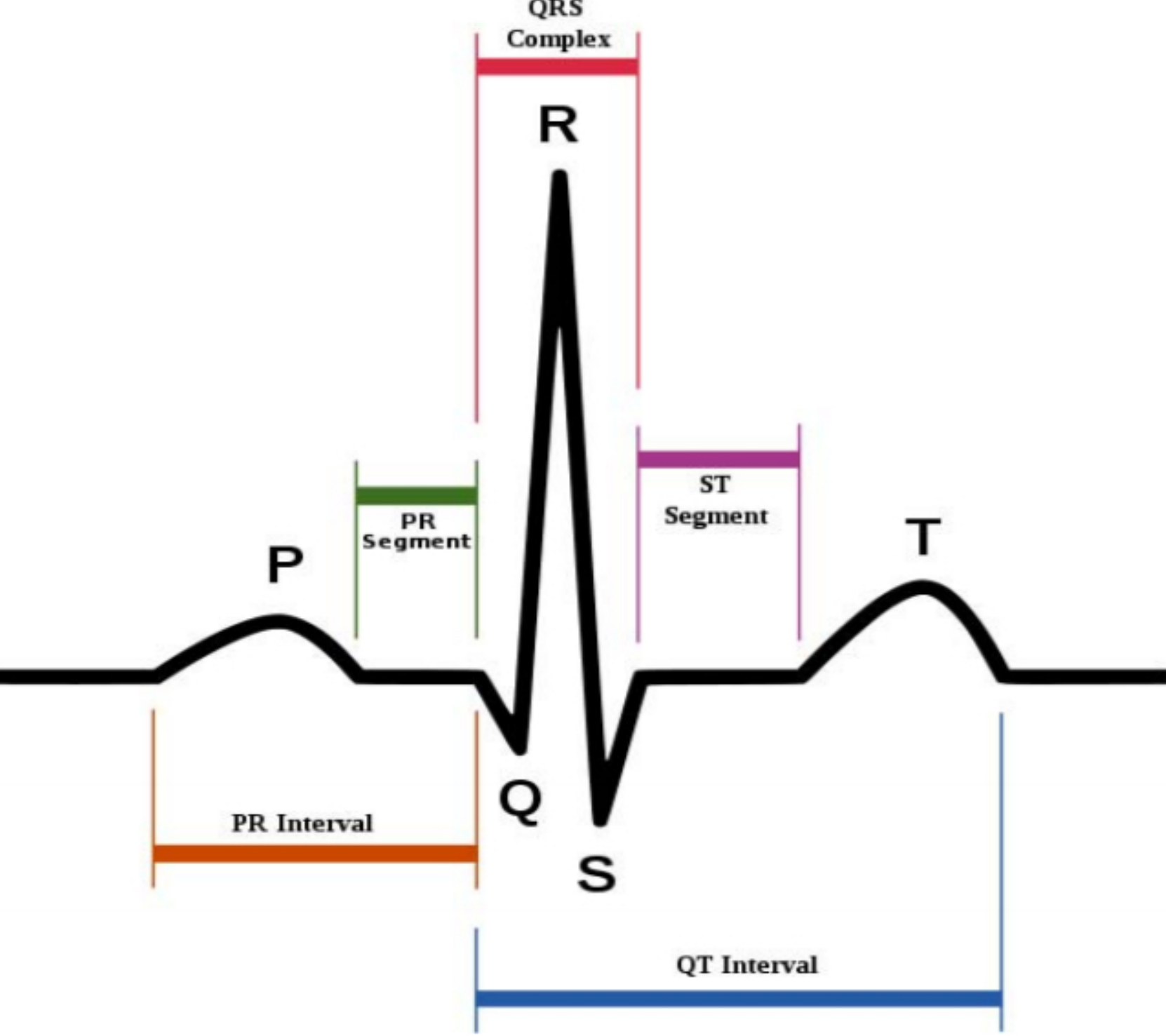}\\
  \caption{A real time ECG signal~\cite{electrocardiograph:2011}.}
  \label{xindiantu}
\end{figure}

Containing abundant information about individual identity, ECG waveform has four fundamental characteristics required for biometric identification~\cite{Minhthang:2008}: (1) Universality: the heart of every living human generates ECG signals at all times; (2) Uniqueness: the ECG differences between individuals are mainly affected by body shape, age, weight, emotion, gender, heart location, heart size, geometric shape, physiological characteristics, chest structure and sports status, etc., so the ecg signals generated by different people are also unique~\cite{Chan:2006}; (3) Stability: the structure and size of the adult heart are basically fixed. ECG waveform remains stable unless the heart has lesions; (4) Measurability: ECG acquisition equipment with miniaturization, portability and high precision reduces the cost and time of ECG acquisition and thus makes the measurement more convenient.
In recent years, ECG signal has become the focus of major research in the field of human identification technology due to its good non-replicability and uniqueness. The methods of ECGID can be divided into two categories: (1) Fiducial approach: as shown in Fig.~\ref{xindiantu}, ECG signal is composed by three main waves the P, the QRS and T waves \cite{electrocardiograph:2011,Cui:2019}, between which the peaks, slopes, boundaries and intervals are depicted by  Fiducial features. The fiducial method utilizes these features for human identification; (2) Non-fiducial approach: these methods often treats ECG signals as a whole without considering the details of waveform, and process signals in the frequency domain in most cases.

Biel et al. extract 30 time and amplitude features and use correlation matrix analysis to select 21 features to form the eigenvector \cite{Biel:2001}. It was demonstrated that soft independent modelling of class analogy method is implemented for classification. Israel et al. locates the key points by finding the local extremum of the waveform in each heart beat and extract 15 features to classify subjects by linear discriminant analysis \cite{Israel:2005}. This experiment included 29 subjects, with the rate of identification reaching 100\% and the rate of heart beat recognition reaching 81\%. Saechia et al. uses Fourier coefficients to examine the effectiveness of segmenting ECG heartbeat \cite{Saechia:2005}. Shen et al. proposed an identification method based on one-lead ECG \cite{Shen:2002}. They obtained 7 time domain features from QRS complex and used template matching and decision-based neural network to complete the verification, with 100\% correct verification rate. Plataniotis et al. first introduced an ECG biometric recognition method requiring no waveform detection and analyzed the auto-correlation of ECGs while applying DCT to achieve dimensionality reduction \cite{Plataniotis:2006}. Wang et al. followed the AC-DCT method and achieved satisfying results \cite{Wang:2008}. Agrafioti et al. combined the auto-correlation and LDA to make the classification \cite{Agrafioti:20081}.
Although these literatures have obtained good results, there remains some important problems that cannot be overlooked, such as robustness of ECGID, including, (1) heart disease: one was enrolled in good heart health, but tested in heart disease; (2) emotion: one was enrolled in calm, but tested in emotional state; (3) exercise: one was enrolled in rest, but tested after exercise(we call it rest-ex ECGID). Some researchers~\cite{Agrafioti:2009,Zhang:20170} have explored the first question with the MIT-BIH database~\cite{MITBIH} or the PTB database~\cite{PTB}, but the latter two issues have rarely been studied. Therefore, we would like to discuss the third issue i.e. the robustness of rest-ex ECGID in this manuscript.

From the 1970s to the 1990s, some researchers\cite{Simmon:1975,Goldberger:1983,Hijzen:1985,Deckers:1990} attempted to dig out the  changes of ECG waveform between rest and exercise, showed that during exercise the amplitude of P wave increases and the amplitude of T wave decreases, while the QRS duration remains nearly constant. After exercise, during the first minute, the amplitude of P and T wave increase and the other features start to return to normal. Then, P and T wave start to return to normal. But they only got qualitative results, not quantitative ones. Afterwards, the literature on the effects of exercise on ECGID also appears, but few good results on rest-ex ECGID have been achieved since post-exercise is a period of recovery where various features change abruptly. Kim et al.\cite{Kyeong:2005} discussed the the effects of exercise on QRS interval, RT interval and QT interval. Furthermore, they used inverse Fourier transform to normalize features and improved rest-ex ECGID performance. Poree et al.\cite{Fabienne:2009} used correlation coefficient for template matching to find out the effects of exercise, waveform length and the number of leads on ECGID. 
Piao et al.\cite{Piao:2016} collected ECG records of only 5 normal people in rest and movement states. They used discrete wavelet transform for feature extraction, Euclidean distance and nearest neighbour for classification, with an accuracy of 100\%. Sung et al.\cite{Sung:2017} collected ECG data of 55 subjects before and after exercise for 5 minutes respectively. The first and second order differential of waveform were used to help feature point detection. LDA is for feature extraction and classification, with the accuracy of identification being 59.64\% within 1 minute after exercise, and more than 90\% beyond 1 minute. Nobunaga et al.\cite{Nobunaga:2017} showed that after bandpass filtering between 10 Hz and 80 Hz, the ECG waveforms in rest and post-exercise would be highly similar, and the rest-ex ECGID rate achieved 99.7\% on 10 subjects. Komeili et al.\cite{Komeili:2016} extracted different types of features and used feature selection algorithms to select the most stable ones in exercise.

This manuscript is aimed at applying existing methods to analyse the performance of rest-ex ECGID with our ECG database \cite{Li:2019}. Through multiple sets of experiments, we find that for current methods, the rest-rest ECGID accuracy can reach more than 95\% while the ex-ex ECGID performance become a little worse; As for the rest-ex ECGID performance, most methods collapse to 10\% or even worse, except for the KL feature selection method, which can reach almost 65\%. Although compared with other methods the KL feature selection method performs much better, its result still remains unsatisfactory. The rest of the paper is organized as follows. Section II introduces the main methods of the existing ECGID algorithms. In Sec. III, the process and results of the experiment with our ECGID database are described in detail. Finally, Sec. IV is the conclusion of the whole manuscript.

\section{Method}

\subsection{Time domain features}
A classical QRS detection algorithm based upon digital analyses of slopes, amplitude, and width, has been proposed by Pan and Tompkins\cite{Pan:1985}, where the ideal pass band intended to maximize the QRS energy is approximately 5-15 Hz. After filtering, the signal is differentiated to provide the QRS complex slope information with the following transfer function
\begin{equation}
H(Z)=\frac{1}{8}[2+Z^{-1}-Z^{-3}-2Z^{-4}].
\end{equation}
Then the signal is squared point by point in order to make positive all of the points and emphasize higher frequencies, given by the relation below
\begin{equation}
X=Y^2.
\end{equation}
In order to obtain waveform feature information, a moving-windows integrator is passed through by squared signals, which is calculated from
\begin{equation}
H(Z)=\frac{1}{N}[Z^{N-1}+Z^{N-2}+\cdots+Z^{N-(N-1)}+1],
\end{equation}
where $N$ is the number of samples in a given width of the integration window.

In the last step, two thresholds are designed to make sure no peak is omitted: the first identifies peaks of the signal and when no peak has been detected by the first one in a certain time interval, the algorithm will search back promptly for the lost peak using the second threshold below the first one.
Once a new peak is identified, it will be classified as a noise peak or as a signal peak if exceeding the first thresholds (or the second threshold if we search back in time for a lost peak). A peak to be identified as a QRS complex must be first identified as a QRS in both the integration and filtered waveforms by investigating the signals and trying different values of the above thresholds.

\subsection{Short-time Fourier transform}
Short-time Fourier transform (STFT)~\cite{Nawab:1983,Griffin:1984} is one of the most commonly used time-frequency analysis method, which represents the signal characteristics at a certain moment through a segment of the signal in the time window. In the short-time Fourier transform process, the length of the window determines the time resolution and frequency resolution of the spectrum. The longer the window is, the higher the frequency resolution is and the lower the time resolution is after the Fourier transform. On the contrary, the shorter the window is, the lower the frequency resolution will be, and the higher the time resolution is. That is to say, in the STFT, the time resolution and frequency resolution cannot be achieved at the same time, and the choice should be made according to specific requirements. Due to the nonstability of ECG signal, STFT can not only transform ECG signal into frequency domain to avoid complex and difficult fiducial feature detection, but also overcome the disadvantage that the original Fourier transform cannot process non-stationary signal. Short-time Fourier transform of the continuous time signal $s(t)$ is
\begin{equation}
\begin{aligned}
STFT(t,\omega) & = \int_{-\infty}^{\infty} s(\tau)\gamma^\ast(\tau-t)e^{-j\omega\tau}\,d\tau \\
& =\langle s(\tau),\gamma(\tau-t)e^{j\omega\tau}\rangle \\
& =\langle s(\tau),\gamma_{t,\omega}(\tau)\rangle
\end{aligned}
\end{equation}
where $\gamma(t)$ is the window function and generally $\gamma(t)$ is real even function.

\subsection{Wavelet transform}
Wavelet transform(WT)~\cite{Daubechies:1990,Addison:2005,LiYang:2018} is a transform analysis method, which inherits and develops the idea of localization of STFT. It overcomes the disadvantages of unchangable window size with frequency, and can provide a "time-frequency" window changing with frequency. It is an ideal tool for signal time-frequency analysis and processing. The main characteristic is that WT can fully highlight some aspects of the problem characteristics,  and can analyse the localization in time (space) frequency. In addition, signal is multi-scale tessellated through the telescopic translation operations, and ultimately adapts to the requirement of time-frequency signal analysis automatically. In other words, WT can focus on the arbitrary signal details, solved the difficulties of Fourier transform. The mathematical expression of the wavelet transform is
\begin{equation}
	WT(a,\tau)=\frac{1}{\sqrt{a}}\int_{-\infty}^{\infty} f(t)*\Psi(\frac{t-\tau}{a})d\tau,
\end{equation}
where $a$ is the scale, corresponding to the frequency and controlling the scaling of the wavelet function  and $\tau$ is the amount of translation, corresponding to the time and controlling the translation of the wavelet function.

\subsection{Autocorrelation}
The ECG is a nonperiodic but highly repetitive signal. The motivation behind the employment of auto correlation based features is to detect the nonrandom patterns. Autocorrelation provides an automatic, shift invariant accumulation of similarity characteristics over multiple heartbeat cycles, which represents the unique characteristics of a given ECG~\cite{Wang:2008}. It provides the most robust representation of the heartbeat characteristics of an subject. In addition, AC is used to blend into a sequence of sums of products samples that would otherwise need to be subjected to fiducial detection. The AC method involves four stages: (1) windowing, where the preprocessed ECG trace is segmented into nonoverlapping windows, with the only restriction that the window has to be longer than the average heartbeat length so that multiple heartbeats are included; (2) estimation of the normalized autocorrelation of each window; (3) feature dimension reductioon, like DCT, LDA PCA etc; and (4) classification.

The autocorrelation coefficients $\widehat{R}_{xx}\left[m \right]$ can be computed as follows
\begin{equation}
    \widehat{R}_{xx}\left[m \right]=\frac{\sum_{i=0}^{N-\vert m\vert -1} {x\left[ i\right] x\left[ i+m\right]}}{\widehat{R}_{xx}\left[0 \right] }
\end{equation}
where $x\left[i \right]$ is the windowed ECG and $x\left[i+m \right]$ is the time-shift version of the windowed ECG with a time lag of $m=0,1,...,M-1, M\ll N$. The division with the maximum value, $\widehat{R}_{xx}\left[0 \right]$, eliminates the bias factor so that either biased or unbiased autocorrelation estimation can be performed. The main contributors to the autocorrelated signal are the P wave, the QRS complex, and the T wave. However, even among the pulses of the same subject, large variations in amplitude present and this makes normalization a necessity. It should be noted that a window is allowed to blindly cut out the ECG record, even in the middle of a pulse. This alone releases the need for exact heartbeat localization.

Autocorrleation provides important information in distinguish subjects. However, the dimension of the autocorrelation coefficient is determined by the size of M, which is generally large(e.g., M=100, 200, 300). So, a dimension reduction technique should be used before classification.

\subsection{Deep learning}
One of the biggest difficulties in ECGID is feature extraction. What features represent a person's ecg characteristics best is a question that we have been looking for a solution to for a long time. But so far, there are no perfect features that fit all situations.  The greatest advantage of deep learning is the ability to learn features automatically, without human intervention.

Deep learning neural networks~\cite{LeCun:2015} is one of the most recent advancements in the machine learning field~\cite{Kiranyaz:2015}. The term refers to multi layer neural networks similar to the ones used in the past. The main difference though is that now it is possible to use more hidden layers than we previously did. In deep learning, because of advancements in hardware and algorithms, there are efficient ways to train these deep neural nets. GPU computing has enabled researchers to utilize more processing power in order to train deep neural nets, while they have also implemented new algorithms that have made the whole procedure more efficient.

Additionally, deep neural nets have been proven successful in image recognition through the use of a new kind of hidden layers, convolutional layers~\cite{Krizhevsky:2012}. These practically enable an hierarchical approach to data input that can detect recurring patterns in different parts of the data inputs. They have been very efficient in image recognition tasks, as they can practically detect objects present in images, irrespectively of the position of the object on the image, or the portion of the image size it corresponds to. Although ECG signal is a one-dimensional signal, it can be transformed into a two-dimensional image by processing in the case of multi-lead, and then be classified by using the convolutional neural network.    Furthermore, Recurrent Neural Network(RNN), invented specifically to classify the time series signals, is another  powerful tool that you can use for ECGID~\cite{Salloum:2017}.

\subsection{Feature selection}
Feature selection approaches can be divided into two categories:
supervised approaches and unsupervised approaches~\cite{Dash:1997}. Supervised approaches use known label information to help select features nevertheless unsupervised approaches focus on mining the internal structure of the data in the absence of label information. In ECGID field, we usually adopt supervised approaches because in the biometric recognition application class labels are provided.

Supervised feature selection methods roughly include filters, wrappers and embedded methods. Wrapper methods, like genetic algorithm~\cite{Yang:1998} and particle swarm optimization, select a feature subset (among all possible feature subsets) that gives the best performance with a specific model. In other words, such algorithms bind to specific models so that the results may be overfitting. In addition, wrappers are computationally very intensive specially if the feature space is too large or the chosen model is complex. Filter methods are relatively fast and do not suffer from aforementioned limitations. Filter method considers each feature individually and designs a criterion to determine whether a feature is useful for classification. Embedded methods embed feature selection in classification.
\section{Experiment}

\subsection{Database}

Our database is collected in Prof. Cui's laboratory at South China University of Technology \cite{Li:2019}.
 The database includes pre and post exercise recordings for 45 subjects. There are 33 males and 12 females whose age is between 18 to 22. Each subject in this set performed a few basic structural workouts such as steady running and climbing the stairs. The length of recordings in rest condition and post-exercise condition are around 5 minutes and 150 seconds respectively. The heart rate in post-exercise condition is range from 90 to 150 compared with around 70 in rest condition. The ECG signal was captured in lead II and the sampling frequency is 300 Hz.

\subsection{Preprocessing}
The raw ECG signals were filtered using a fourth-order bandpass
Butterworth filter with cut-off frequencies 0.5 Hz and 40 Hz~\cite{Wang:2008}. Under 0.5 Hz the signal is corrupted by baseline wander, and over 40 Hz there is distortion due to muscle movement,
power-line noise etc.

\subsection{Experiment on QRS complex}
As mentioned before, QRS complex is the most stable part between rest and post-exercise. We use Pan \& Tompkins algorithm to locate QRS complex. The length of QRS complex varies between 15 and 35 samples, so we normalize the duration of QRS complex to the same periodic length according to the procedure reported in~\cite{Wei:2001}. That is, one of the ECG segments \bm{$y_i$}=$[y_i(1),y_i(2),\cdots,y_i(n^*)] $ can be converted into a segment \bm{$x_i$}=$[x_i(1),x_i(2),\cdots,x_i(n)] $ that holds the same signal morphology, but in different data length (i.e., $n^*\not=n $) using the following equation,
\begin{eqnarray}
 x_i(j)=y_i(j^*)+(y_i(j^*+1)-y_i(j^*))(r_j-j^*),
\end{eqnarray}
 where $r_j=(j-1)(n^*-1)/(n-1)+1 $, and $ j^*$ is the integral part of $r_j $. Here , we set $n=30$. Therefore, the various lengths of the QRS complex will be compressed or extended into a set of ECG segments with the same periodic length. These 30 samples are considered to be the main description of a heart beat, and thus are regarded as the features of the heart beat. Regarding the classification method, we choose RBF kernel with $c=100$ and $\gamma=1$ for SVM after reducing the feature dimension with PCA.
In this manuscript, we mainly perform four types of ECGID test: (1) the training set and test set are both from rest condition~(rest ECGID); (2) the training set and test set are both from post-exercise condition~(ex-ex ECGID). According to the mode of distribution, this type is divided into two subtypes, which are the first 70\% as the training set and the last 70\% as the training set; (3) the training set is from rest condition and the test set is from post-exercise condition~(rest-ex ECGID).
 \begin{center}
 \scriptsize
 { Table~1\\ ECGID performance on QRS complex}\\
 \label{tab:1} \vskip 3pt
 \begin{tabular*}{\linewidth}{p{2cm}p{2cm}p{1.5cm}p{1.5cm}}
  \toprule
Training set & Test set & Training accuracy & Test accuracy  \\
  \midrule
rest record  & rest record & 98\% & 95\%  \\
post-exercise(first 70\%)  & post-exercise(last 30\%)  & 94.1\% & 82.9\%  \\
post-exercise(last 70\%)   & post-exercise(first 30\%)  & 94.8\% & 70.2\%  \\
rest & post-exercise  & 97.9\% & 17.5\%  \\
  \bottomrule
 \end{tabular*}
\end{center}

\subsection{Experiment on ECG beat}
Pan \& Tompkins algorithm is also utilized to locate R peak. Then we search the midpoint of two R-peak samples, and the signal between two consecutive midpoints is defined as a ECG segment. Like the way we used in the experiment on QRS complex, we also use the procedure reported in Ref.~\cite{Wei:2001} to normalize the length of  ECG segments. We set $n=300$ here.  These 300 samples are regarded as the features of the heart beat. We adopted SVM and deeplearning for classification respectively. For deep learning, the network structure we adopt is as shown in Fig.~\ref{LSTM}.
 And the results of experiment on ECG beat are shown in Tables 2-3:

\begin{figure}
  \centering
  \includegraphics[width=2.8cm]{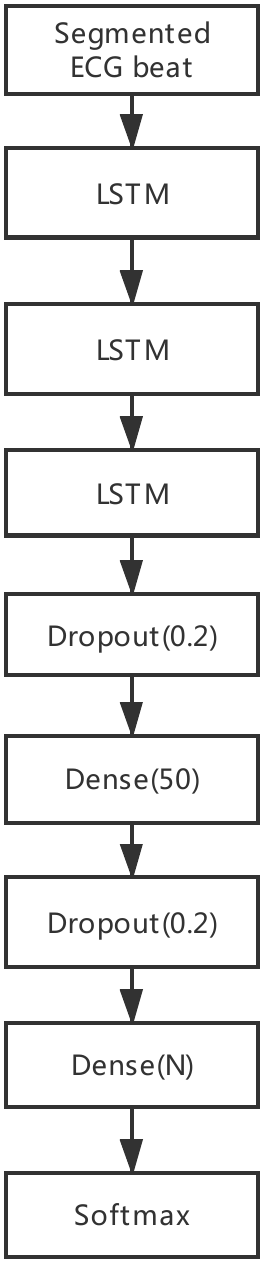}\\
  \caption{Network structure of LSTM on ECGID}
  \label{LSTM}
\end{figure}

 \begin{center}
 \scriptsize
 { Table~2\\ ECGID performance on ECG beat with SVM}\\
\label{tab:2} \vskip 3pt
\newcommand{\rb}[1]{\raisebox{1.9ex}[-2pt]{#1}}
 \begin{tabular*}{\linewidth}{p{2cm}p{2cm}p{1.5cm}p{1.5cm}}
  \toprule
Training set & Test set & Training accuracy & Test accuracy  \\
  \midrule
rest record  & rest record & 100\% & 96.2\%  \\
post-exercise(first 70\%)  & post-exercise(last 30\%)  & 100\% & 85.4\%  \\
post-exercise(last 70\%)   & post-exercise(first 30\%)  & 100\% & 72.6\%  \\
rest   & post-exercise  & 99.6\% & 2.7\%  \\
  \bottomrule
 \end{tabular*}
\end{center}

 \begin{center}
 \scriptsize
 { Table~3\\ ECGID performance on ECG beat with LSTM}\\
\label{tab:3} \vskip 3pt
\newcommand{\rb}[1]{\raisebox{1.9ex}[-2pt]{#1}}
 \begin{tabular*}{\linewidth}{p{2cm}p{2cm}p{1.5cm}p{1.5cm}}
  \toprule
Training set & Test set & Training accuracy & Test accuracy \\
  \midrule
rest record  & rest record & 97.4\% & 95.6\%  \\
post-exercise(first 70\%)  & post-exercise(last 30\%)  & 96\% & 82.3\%  \\
post-exercise(last 70\%)   & post-exercise(first 30\%)  & 97.2\% & 77.6\%  \\
rest   & post-exercise  & 98.2\% & 12\%  \\
  \bottomrule
 \end{tabular*}
\end{center}

\subsection{Experiment on PQRST piecewise correction}
In this experiment, we divide a complete heart beat into four parts: PQ segment, QRS segment, ST segment and T segment. Changes in heart rate are not evenly distributed across the P, R, and T complexes, so we adopt piecewise correction on these segments as follows. The PQ segment can be obtained using the following Equation 
\begin{equation}
 PQ(i)=S[R(i)-230ms+dt:R(i)-90ms],
\end{equation}
where $dt$ is a variable threshold with changes in heart rate,
and it can be described as follows
\begin{equation}
dt=\left\{
 		\begin{array}{lr}
        -10ms,  ~~~~~~HR<65 & \\
        0ms,  ~~~~~~65\leq HR<80 & \\
        10ms,  ~~~~~80\leq HR<95 & \\
        20ms,  ~~~~~95\leq HR<110 & \\
        30ms,  ~~~~~110\leq HR<125 & \\
        40ms,  ~~~~~125\leq HR<140 & \\
        50ms,  ~~~~~140\leq HR<155 &
        \end{array}.
\right.
\end{equation}
The QRS-segment can be obtained using Equation (10)
\begin{equation}
 QRS(i)=S[R(i)-90ms:R(i)+100ms].
\end{equation}
The ST-segment can be obtained using Equation (11)
\begin{equation}
 ST(i)=S[R(i)+100ms:R(i)+100ms+0.08*RR],
\end{equation}
where the $RR$ is current RR interval. As for the T-segment, it is described by Equation (12)
\begin{equation}
 T(i)=S[R(i)+100ms+0.08*RR:R(i)+0.42*RR].
\end{equation}
After wave correction, we use the method mentioned above to resample the PQ segment to 450 milliseconds length, the ST segment to 110 milliseconds length, and the T segment to 50 milliseconds length. The QRS segment keeps unchanged for it remains fairly constant. These segments are combined to recreate the entire heartbeat. The normal length of a heartbeat is 800 milliseconds. Finally, the amplitude of each new heartbeat is also normalized to a mean zero. Next, we perform two kinds of processing. One is to  classify the new heart beat as a feature vector through PCA and SVM, directly; the other is to extract the wavelet coefficient with the new heart beat and then conduct identity classification. The experiment result is shown in  Table 4: 

 \begin{center}
 \scriptsize
 { Table~4\\ ECGID performance on PQRST piecewise correction}\\
\label{tab:4} \vskip 3pt
\newcommand{\rb}[1]{\raisebox{1.9ex}[-2pt]{#1}}
 \begin{tabular*}{\linewidth}{p{2cm}p{2cm}p{1.5cm}p{1.5cm}}
  \toprule
Training set & Test set & Training accuracy & Test accuracy \\
  \midrule
rest record  & rest record & 100\% & 98.2\%  \\
post-exercise(first 70\%)  & post-exercise(last 30\%)  & 100\% & 88\%  \\
post-exercise(last 70\%)   & post-exercise(first 30\%)  & 100\% & 63.3\%  \\
rest   & post-exercise  & 99.9\% & 4.9\%  \\
  \bottomrule
 \end{tabular*}
\end{center}


\subsection{Experiment on band pass filtering}
Ref.\cite{Nobunaga:2017} showed that the ECG exercise waveform is similar to the rest waveform above 10 Hz, so in this experiment, after R peak detecting and ECG heartbeat segmentation, we filter the original ECG signal, including rest and post-exercise, from 10 Hz to 40 Hz. Then the heart beats are resampled to 300 points length and a rbf-SVM is used for classification. The experiment result is as follows (see Table 5):
 \begin{center}
 \scriptsize
 { Table~5\\ ECGID performance on band pass filtering}\\
\label{tab:5} \vskip 3pt
\newcommand{\rb}[1]{\raisebox{1.9ex}[-2pt]{#1}}
 \begin{tabular*}{\linewidth}{p{2cm}p{2cm}p{1.5cm}p{1.5cm}}
  \toprule
Training set & Test set & Training accuracy & Test accuracy \\
  \midrule
rest   & rest  & 100\% & 92.7\%  \\
post-exercise(first 70\%)  & post-exercise(last 30\%)  & 100\% & 85.4\%  \\
post-exercise(last 70\%)   & post-exercise(first 30\%)  & 100\% & 52.4\%  \\
rest   & post-exercise  & 98.6\% & 3\%  \\
  \bottomrule
 \end{tabular*}
\end{center}

\subsection{Experiment on Short-time Fourier transform}
In this experiment, we use short-time Fourier transform with Hamming
window of the length 16 with step size of 13 computed over a 1 second window centered at R peak. The sampling frequency is 300 Hz, so this gives a total of 572 features. A PCA is for dimension reduction and a rbf-SVM is for classification. The experiment result  is as follows (see Table 6):
 \begin{center}
 \scriptsize
 { Table~6\\ ECGID performance on STFT}\\
\label{tab:6} \vskip 3pt
\newcommand{\rb}[1]{\raisebox{1.9ex}[-2pt]{#1}}
 \begin{tabular*}{\linewidth}{p{2cm}p{2cm}p{1.5cm}p{1.5cm}}
  \toprule
Training set & Test set & Training accuracy & Test accuracy \\
  \midrule
rest   & rest  & 99.6\% & 98.3\%  \\
post-exercise(first 70\%)  & post-exercise(last 30\%)  & 97.1\% & 75.3\%  \\
post-exercise(last 70\%)  & post-exercise(first 30\%)  & 98.3\% & 66.1\%  \\
rest   & post-exercise  & 99.6\% & 12.9\%  \\
  \bottomrule
 \end{tabular*}
\end{center}

\subsection{Experiment on Wavelet transform}
After preprocessing, Continuous wavelet transform with 32 scales and Daubechies 5 as mother wavelet is computed on a 1 second window centered at R peak, which gives a total of 9600 features. A PCA is for dimension reduction and a rbf-SVM is for classification. The experiment result is as follows (see Table 7):
 \begin{center}
 \scriptsize
 { Table~7\\ ECGID performance on wavelet transform}\\
\label{tab:7} \vskip 3pt
\newcommand{\rb}[1]{\raisebox{1.9ex}[-2pt]{#1}}
 \begin{tabular*}{\linewidth}{p{2cm}p{2cm}p{1.5cm}p{1.5cm}}
  \toprule
Training set & Test set & Training accuracy & Test accuracy \\
  \midrule
rest   & rest  & 99.9\% & 99.7\%  \\
post-exercise(first 70\%)  & post-exercise(last 30\%)  & 99.8\% & 83\%  \\
post-exercise(last 70\%)  & post-exercise(first 30\%)  & 99.9\% & 77.9\%  \\
rest   & post-exercise  & 99.9\% & 7.1\%  \\
  \bottomrule
 \end{tabular*}
\end{center}

\subsection{Experiment on Autocorrelation}
Autocorrelation upto $n$ lags is computed on windows of the
length $L$ seconds, where $n$ and $L$ are adjustable parameters. The experiment results are shown in Tables 8 \&  9:
 \begin{center}
 \scriptsize
 { Table~8\\ ECGID performance on Autocorrelation}\\
\label{tab:8} \vskip 3pt
\newcommand{\rb}[1]{\raisebox{1.9ex}[-2pt]{#1}}
 \begin{tabular*}{\linewidth}{p{2cm}p{2cm}p{1.5cm}p{1.5cm}}
  \toprule
Training set & Test set & Training accuracy & Test accuracy \\
  \midrule
rest   & rest  & 100\% & 93.8\%  \\
post-exercise(first 70\%)  & post-exercise(last 30\%)  & 99.9\% & 74.3\%  \\
post-exercise(last 70\%)   & post-exercise(first 30\%)  & 100\% & 60.1\%  \\
rest   & post-exercise  & 90\% & 11.3\%  \\
  \bottomrule
 \end{tabular*}
\end{center}

 \begin{center}
 \scriptsize
 { Table~9\\ ECGID performance on beat with Autocorrelation}\\
\label{tab:9} \vskip 3pt
\newcommand{\rb}[1]{\raisebox{1.9ex}[-2pt]{#1}}
 \begin{tabular*}{\linewidth}{p{2cm}p{2cm}p{1.5cm}p{1.5cm}}
  \toprule
Training set & Test set & Training accuracy & Test accuracy\\
  \midrule
rest   & rest  & 100\% & 95.2\%  \\
post-exercise(first 70\%)  & post-exercise(last 30\%)  & 99.2\% & 67.3\%  \\
post-exercise(last 70\%)   & post-exercise(first 30\%)  & 99.6\% & 63.9\%  \\
rest   & post-exercise  & 98.3\% & 1.3\%  \\
  \bottomrule
 \end{tabular*}
\end{center}

\subsection{Experiment on feature selection}
We randomly select half of the subjects (22 subjects) as an auxiliary data set and utilize the other half (23 subjects) for registration and testing. In other word, the first half is picked for feature selection while the second one for calculating accuracy of identification where registration is performed in the resting state and testing in the post-exercise condition.
We perform a short-time Fourier transform using a Hamming window of length 16 seconds with a step size of 13 to, and compute continuous wavelet transform with 32 scales and Daubechies 5 as mother wavelet over a window of length 1 second centered at R peak. Autocorrelation up to 80 lags is calculated over 1 second windows. In order to have zero mean and unit variance, every feature has to be normalized by z-score in the feature set which is collected by 10252 features.  

\begin{figure}
  \centering
  \includegraphics[width=8cm]{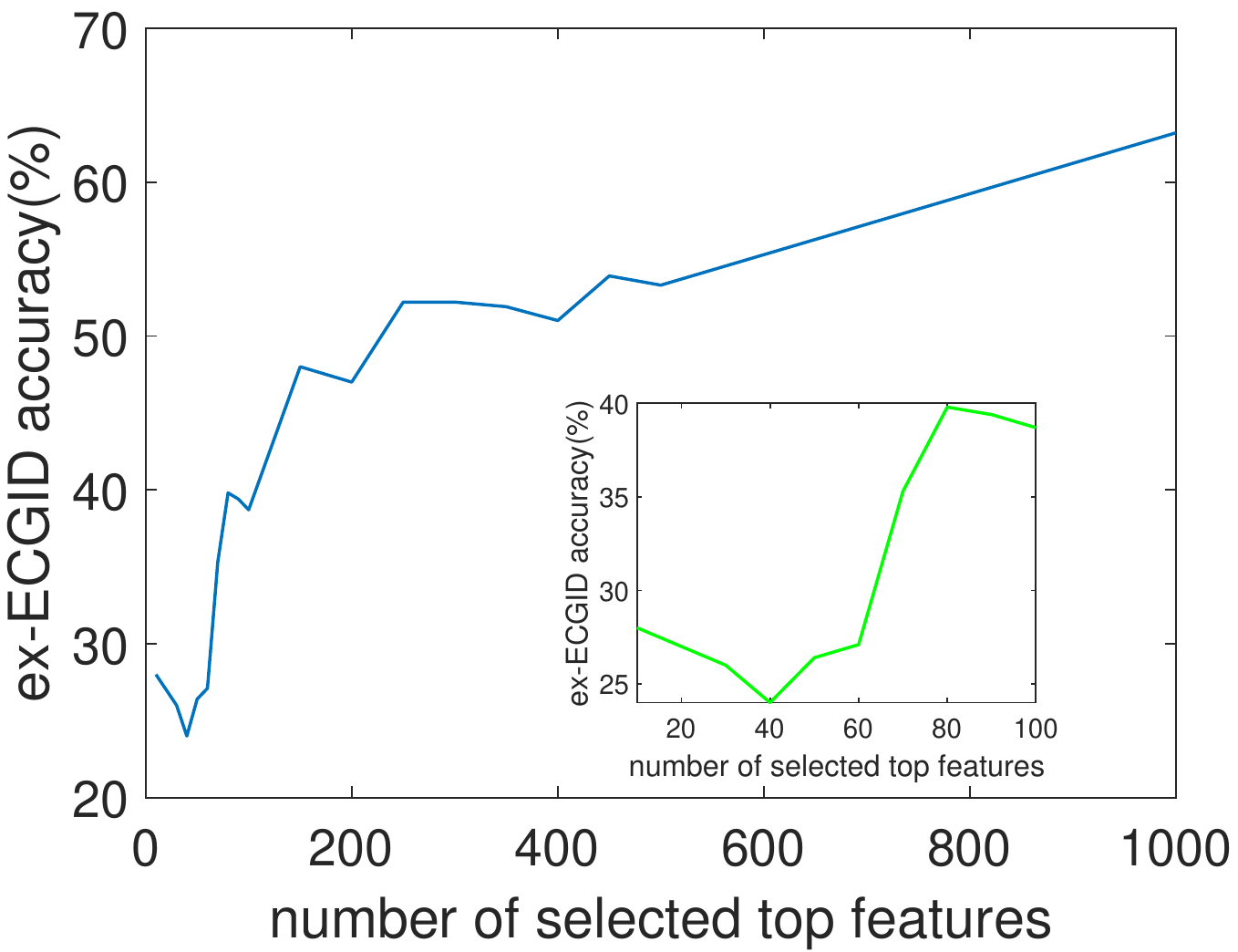}\\
  \caption{Rest-ex ECGID accuracy of KL feature selection}
  \label{KL}
\end{figure}

The feature selection is based on Kullback-Leibler (KL) divergence. Following Ref.\cite{Komeili:2016}, feature weights, $w(l)$, which measure the importance of a feature to rest-ex ECGID, are defined as linear combination of two terms
\begin{equation}
w(l)=\lambda w_1(l)-(1-\lambda)w_2(l).
\end{equation}
The first term $w_1(l)$ related to the class separability is
defined as follows
\begin{equation}
w_1(l)=\frac{1}{N}\sum_{i=1}^Nd(f(X_i(l)), f(\chi(l))),
\end{equation}
where $f(X_i(l))$ is probability density function(pdf) of $l$-th feature computed over all samples of $i$-th subject including both rest and post-exercise samples. $f(\chi(l))$ is pdf of $l$-th feature computed over all samples of the auxiliary data set while $N$ is total number of its subjects. $d(\cdot)$ is the symmetric KL divergence which can be estimated under a normal distribution as follows
\begin{equation}
d(f_1,f_2)=\frac{\sigma_1^2+(\mu_1-\mu_2)^2}{2\sigma_2^2} + \frac{\sigma_2^2+(\mu_1-\mu_2)^2}{2\sigma_1^2}-1.
\end{equation}
The second term in equation (13) related to the sensitivity of a feature to exercise is defined below
\begin{equation}
\begin{aligned}
w_2(l)=\frac{1}{N}\sum_{i=1}^N(d(f(X_i^{rest}(l)), f(X_i(l))) & \\
+ d(f(X_i^{post-ex}(l)), f(X_i(l)))),
\end{aligned}
\end{equation}
where $f(X_i^{rest}(l))$($f(X_i^{post-ex}(l))$) is part of $l$-th feature of subject $i$ in rest(post-exercise) condition. $w_2(l)$ is small when the distribution corresponding to rest and post-exercise condition overlap with each other, suggesting the feature is more robust against exercise. Features can be selected by comparing the weights with a threshold and further sorted in descending order according to their weights, after which the top $n$ features are picked. $\lambda$ is chosen to be 0.3 empirically. The ECGID performance on KL feature selection is shown in Table 10, while the details of rest-ex ECGID performance is shown in Fig.~\ref{KL}.

 \begin{center}
 \scriptsize
 { Table~10\\ ECGID performance on KL feature selection}\\
\label{tab:10} \vskip 3pt
\newcommand{\rb}[1]{\raisebox{1.9ex}[-2pt]{#1}}
 \begin{tabular*}{\linewidth}{p{2cm}p{2cm}p{1.5cm}p{1.5cm}}
  \toprule
Training set & Test set & Training accuracy & Test accuracy \\
  \midrule
rest   & rest  & 98.2\% & 93.8\%  \\
post-exercise(first 70\%)  & post-exercise(last 30\%)  & 95.7\% & 85.1\%  \\
post-exercise(last 70\%)   & post-exercise(first 30\%)  & 96.4\% & 73.9\%  \\
rest   & post-exercise  & 98.7\% & 61.4\%  \\
  \bottomrule
 \end{tabular*}
\end{center}

It is shown that compared with other methods, though the KL feature selection method can significantly improve the rest-ex ECGID accuracy, it is only 61.4\%, far from satisfying the expectation.

\section{Conclusion}
In this manuscript, we focus on the effect of exercise on ECGID and apply existing methods to analyse the performance of rest-ex ECGID with our ECG database. Through multiple sets of experiments, we find that for current methods, the rest-rest ECGID accuracy can reach more than 95\% while the ex-ex ECGID performs a little worse; As for the rest-ex ECGID performance, most methods collapse to 10\% or even worse, except for the KL feature selection method which can reach almost 65\%. Although compared with other methods the KL feature selection method performs much better, its result still remains unsatisfactory. The current rest-ex ECGID solutions are too simple, but more stable features may be obtained and sent to feature selection algorithm in the future by filtering out the part with violent changes due to exercise. Furthermore, other more advanced methods in the field of identity recognition can be introduced to improve the accuracy of rest-ex ECGID.

\vskip 12pt
 {\fontsize{7.8pt}{9.4pt}\selectfont
}
%
%

\end{document}